# Revision of Sabine's reverberation theory by following a different approach to Eyring's theory

Toshiki Hanyu [1*]

[1] *Nihon University, Junior College, Dept. of Architecture and Living Design, 7-24-1 Narashinodai, Funabashi, Chiba, 274-8501 Japan*

**Abstract:** The room acoustic theory was established based on Sabine's reverberation theory. However, in Sabine's theory, the reverberation time does not reach zero, even if the absolute absorption condition is satisfied. This is a contradiction of Sabine's theory, and Eyring revised the reverberation theory to resolve this contradiction. In this paper, a theoretical framework for the consistent reverberation theory is presented. Using this framework, it was demonstrated that Eyring's theory has a contradiction between the sound energy density in the steady state and energy decay from the steady state, which is absent in Sabine's theory. Based on the proposed theoretical framework, Sabine's reverberation theory was revised using an approach that is different from that of Eyring. The reverberation time obtained using the revised theory was shorter than that obtained using Sabine's theory and longer than that obtained using Eyring's theory. Results of sound ray tracing simulations were in better agreement with the values calculated using the revised theory rather than those calculated using Sabine's and Eyring's theories.

**Keywords:** Reverberation theory, Reverberation time, Sabine, Eyring, Revised theory

## 1. INTRODUCTION

The room acoustic theory is based on Sabine's [1] and Eyring's [2] reverberation theories. Over the years, it has been improved from various points of view, such as the concept of averaging absorption [3], air absorption [4], directional reverberation [5,6], uneven absorption [7,8], and consideration of sound scattering [9–11]. However, even in present times, Sabine's and Eyring's theories are the most important theoretical frameworks for room acoustics. The important physical quantities derived from these theories are the average sound pressure level (sound energy density) and reverberation time (sound energy decay). Despite that, Sabine's and Eyring's theories are incompatible in these aspects, and are not an integrated and consistent theoretical system. In particular, Eyring's theory has contradictions regarding the sound energy density in a steady state and energy decay from the steady state.

The author has been studying to construct a consistent room acoustic theoretical system [12,13]. This study presents the theoretical framework for the consistent system, and the contradictions in the current room acoustic theory are clarified. It aims to revise Sabine's reverberation theory by following a different approach to Eyring's theory based on the proposed theoretical framework.

## 2. FRAMEWORK OF REVERBERATION THEORY

### 2.1. Theoretical framework

As in Sabine's and Eyring's theory, a diffuse sound field is assumed in this theoretical framework. To meet this assumption, a room must not have extreme dimensions or uneven absorption distribution. It is also assumed that reflections from a wall are totally diffuse and follow Lambert's cosine law.

It is assumed that a sound source with sound power $W$ emits sound for an extremely short time $\Delta t$ in a room with volume $V$ and that the sound energy density $E$ exponentially decays with a decay parameter $\lambda$. This is expressed using Eq. (1). This equation expresses the average impulse response in the room.

$$\Delta E(t) = \frac{W \Delta t}{V} \exp(-\lambda t). \qquad (1)$$

Eq. (2) is obtained by setting $\Delta t \to 0$.

$$\frac{dE(t)}{dt} = \frac{W}{V} \exp(-\lambda t). \qquad (2)$$

* hanyu.toshiki@nihon-u.ac.jp

The Schroeder integration [14] of Eq. (2) yields the following equation, which expresses the sound energy decay from the steady state of the sound fields:

$$E(t) = \frac{W}{V} \int_t^\infty \exp(-\lambda \tau) \, d\tau = \frac{W}{V} \frac{1}{\lambda} \exp(-\lambda t). \quad (3)$$

From Eq. (3), the sound energy density $E_0$ in the steady state can be defined as Eq. (4) using $\lambda$. Eq. (3) can be rewritten as Eq. (5) by using $E_0$. In this theoretical framework, $\exp(-\lambda t)$ is construed as the time variation of the probability that the emitted sound energy remains in a room.

$$E_0 = \frac{W}{V\lambda}. \quad (4)$$

$$E(t) = E_0 \exp(-\lambda t). \quad (5)$$

In the theoretical framework, the sound energy density of the steady state and reverberation decay are determined by the identical parameter $\lambda$. Parameter $\lambda$ is a key factor in the room acoustic theory. Thus, to achieve a consistent theoretical system, parameter $\lambda$ in the sound energy density $E_0$ and the reverberation decay $\exp(-\lambda t)$ should be identical.

**2.2. Reverberation time and average sound pressure level in the theoretical framework**

The reverberation time and average sound pressure level are defined using $\lambda$ according to the theoretical framework. The reverberation time $T$ is defined as the time in which the sound energy density becomes $10^{-6} E_0$. This is expressed as follows:

$$10^{-6} = \exp(-\lambda T). \quad (6)$$

Solving for $T$ defines the reverberation time using $\lambda$ as follows:

$$T = \frac{6 \ln(10)}{\lambda} \quad [\text{s}] \quad (7)$$

Using $I = cE$, Eq. (4) can be transformed into $I = cW/V\lambda$ where $I$ is the sound intensity, and $c$ is the speed of sound. Therefore, when the sound power level is $L_w$, the average sound pressure level $L_p$ can be calculated using Eq. (8) as follows:

$$L_p = L_w + 10 \log_{10}\left(\frac{c}{V\lambda}\right) \quad [\text{dB}]. \quad (8)$$

As shown in Eqs. (7) and (8), important physical quantities in room acoustics are determined using the identical parameter $\lambda$.

## 3. THEORIES OF SABINE AND EYRING REGARDING THE FRAMEWORK

### 3.1. Sabine's theory

In Sabine's theory, reverberation decay from the steady state can be expressed using Eq. (9), where $S$ and $\bar{\alpha}$ are the total surface area and average absorption coefficient of the room, respectively.

$$E(t) = \frac{W}{V} \frac{4V}{cS\bar{\alpha}} \exp\left(-\frac{cS\bar{\alpha}}{4V} t\right). \quad (9)$$

From Eq. (9), $\lambda$ becomes $cS\bar{\alpha}/4V$ identically both in the sound energy density of the steady state and in the reverberation decay. Therefore, Sabine's theory is consistent with the theoretical framework. If $\lambda = cS\bar{\alpha}/4V$ is substituted into Eqs. (7) and (8), the well-known definitions of $T$ and $L_p$ can be obtained using Eqs. (10) and (11), respectively, where $A$ is an equivalent absorption area.

$$T = \frac{24 \ln(10) V}{cS\bar{\alpha}} = \frac{0.163 V}{A} \quad [\text{s}]. \quad (10)$$

$$L_p = L_w + 10 \log_{10}\left(\frac{4}{A}\right) \quad [\text{dB}]. \quad (11)$$

As mentioned previously, Sabine's theory is consistent with the theoretical framework described in Section 2. However, the reverberation time in Sabine's theory does not reach zero, even if $\bar{\alpha} = 1.0$. This has been considered as a shortcoming of Sabine's theory.

### 3.2. Eyring's theory

In Eyring's theory, the reverberation decay from the steady state becomes $E(n) = E_0 (1 - \bar{\alpha})^n$ where $n$ is an order of reflection. When the mean free path of room $\bar{\ell}$ is defined as $\bar{\ell} = 4V/S$ [15–18], the average reflection order up to time $t$ $\bar{n}$ is $cSt/4V$. Using these relations, the decay $(1 - \bar{\alpha})^{\bar{n}}$ and sound energy density of the steady state $E_0$ can be expressed using Eqs. (12) and (13), respectively, as follows:

$$(1 - \bar{\alpha})^{\bar{n}} = \exp\left[cS \ln(1 - \bar{\alpha}) t / 4V\right], \quad (12)$$

$$E_0 = \frac{W}{V} \frac{\bar{\ell}}{c} \left[1 + \sum_{n=1}^{\infty} (1 - \bar{\alpha})^n\right] = \frac{W}{V} \frac{4V}{cS\bar{\alpha}}. \quad (13)$$

Therefore, according to Eyring's theory, the reverberation decay from the steady state can be expressed using Eq. (14) as follows:

$$E(t) = \frac{W}{V} \frac{4V}{cS\bar{\alpha}} \exp\left[\frac{cS \ln(1 - \bar{\alpha})}{4V} t\right]. \quad (14)$$

Revision of Sabine's Reverberation Theory

From Eq. (14), $\lambda$ in the reverberation decay becomes $-cS\ln(1-\bar{\alpha})/4V$. After substituting $\lambda = -cS\ln(1-\bar{\alpha})/4V$ into Eq. (7), Eyring's reverberation time, can be obtained as follows:

$$T = \frac{24\ln(10)V}{-cS\ln(1-\bar{\alpha})} = \frac{0.163V}{-S\ln(1-\bar{\alpha})} \quad [\text{s}]. \quad (15)$$

From Eqs. (13) and (14), $\lambda$ in $E_0$ becomes $cS\bar{\alpha}/4V$ similar to that in Sabine's theory. Therefore, $L_p$ was defined using Eq. (11). According to Eyring's theory, the reverberation time is zero when $\bar{\alpha} = 1.0$. Thus, the contradiction in Sabine's theory is resolved. However, from Eq. (14), $\lambda$ in $E_0$ and the reverberation decay is different. Therefore, Eyring's theory is inconsistent with the theoretical framework described in Section 2.

## 4. REVISED THEORY

### 4.1. Parameter $\lambda$ in revised theory

Here, we examine a revision of Sabine's theory. In the revised theory, reverberation is defined as "a decay of average sound energy density in the entire space, which includes both direct sound and reflected sounds." Moreover, it was assumed that a perfect diffusion state would be maintained both in a steady state and in the entire reverberation process.

Fig. 1 shows a conceptual diagram of the reverberation decay based on Eq. (2). In Sabine's theory, the reverberation decay in Eq. (2) becomes $(W/V)\exp(-ct/\bar{\ell})$ for $\bar{\alpha}=1$, where $\bar{\ell} = 4V/S$. Because reflected sound is not generated at all when $\bar{\alpha}=1$, this decay has been considered as a contradiction of Sabine's theory. However, this decay is not a contradiction completely in the revised theory.

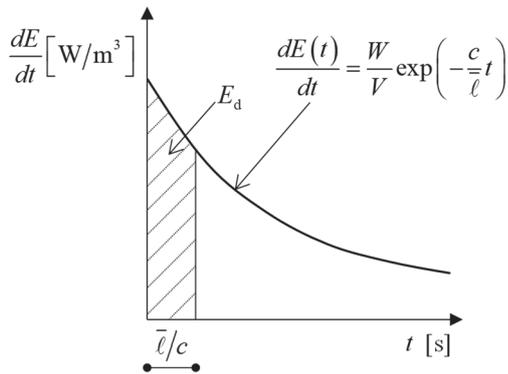

**Fig. 1** Conceptual diagram of the energy density of direct sound.

In the revised theory, this decay is construed as reverberation of direct sound. Basically, a point sound source at an arbitrary point in a room is assumed here. And the direct sound is defined as the sound energy that has not hit wall surfaces at all. The sound energy of the direct sound does not vanish till it reaches wall surfaces even when $\bar{\alpha}=1$. Thus, if the sound source stops in a steady state, the sound energy of the direct sound remains in space for a while. This is the concept of reverberation of direct sound. According to this concept, the reverberation time does not necessarily reach zero, as per Eyring's theory. However, it is also incorrect that the direct sound energy remains far beyond time $t = \bar{\ell}/c$, as per Sabine's theory. In the revised theory, this problem of the reverberation of direct sound is resolved as below.

The revised theory is based on the stochastic concept and describes the expected value of the effect of a certain sound source under the condition of perfect diffusion both in the steady state and in the entire reverberation process. This is equivalent to describing the average effect of numerous point sound sources uniformly distributed throughout the room.

When the numerous sound sources stop simultaneously in a steady state, the average energy density of all direct sounds gradually decreases as the direct sounds continue to hit walls. After eventually all direct sounds hit the walls, the energy of the direct sounds disappears completely.

Here, to resolve the problem of the reverberation of direct sound in Sabine's theory, the maximum time that the direct sounds can remain $t_{d\_max}$ should be examined. First, a free path of direct sound $\ell_d$ is defined as the distance the sound travels from a sound source to a wall. If the maximum length of $\ell_d$ of the numerous sound sources is defined as $\ell_{d\_max}$, $t_{d\_max}$ can be expressed as $\ell_{d\_max}/c$. However, $\ell_{d\_max}$ cannot be determined without a specific room shape. Therefore, based on the stochastic concept, we need to utilize not $\ell_{d\_max}$ but its expected value $\overline{\ell_{d\_max}}$. In that case, the expected value of $t_{d\_max}$, $\overline{t_{d\_max}}$, can be derived as $\overline{\ell_{d\_max}}/c$. $\overline{\ell_{d\_max}}$ can be deduced assuming the extreme condition that numerous sound sources are located on all of the wall surfaces. In this condition, it can be considered that $\overline{\ell_{d\_max}}$ equals to the mean free path $\bar{\ell}$. Therefore, the direct sound is limited to the time $\overline{t_{d\_max}} = \bar{\ell}/c$ on average in the revised theory. Based on this, the energy density of the direct sound $E_d$ can be calculated using Eq. (16).

$$E_d = \frac{W}{V}\int_0^{\bar{\ell}/c} \exp\left(-\frac{c}{\bar{\ell}}\tau\right)d\tau = \frac{W}{V}\frac{\bar{\ell}}{c}(1-e^{-1}). \quad (16)$$

The energy density of the direct sound $E_d$ becomes $W\bar{\ell}/cV$ in both Sabine's and Eyring's theories but $W\bar{\ell}(1-e^{-1})/cV$ in the revised theory.

Eq. (16) contains only the direct sound because Eq. (16) is derived from Eq. (2) with the conditions $\lambda = cS\bar{\alpha}/4V$ and $\bar{\alpha} = 1$. Thus, $E_d$ can be calculated by setting $\bar{\ell}/c$ as the integral interval. However, if $0 < \bar{\alpha} < 1$, direct and reflected sound energy always coexist ($0<t$). Thus, a temporal border does not exist to clearly distinguish the direct and reflected sounds. On the other hand, in Eq. (13), since the energy density in the steady state $E_0$ is defined by using reflection orders $n$ but not time $t$, the direct sound ($n=0$) and reflected sounds ($0<n$) can be clearly distinguished. This is one of the advantages in Eyring's theory. Therefore, in the revised theory, $E_0$ is redefined according to Eyring's manner using $E_d$ calculated by Eq. (16). Based on the above discussions, $E_0$ can be obtained using Eq. (17).

$$E_0 = \frac{W}{V}\frac{\bar{\ell}}{c}\left[(1-e^{-1}) + \sum_{n=1}^{\infty}(1-\bar{\alpha})^n\right]$$
$$= \frac{W}{V}\frac{4V(1-e^{-1}\bar{\alpha})}{cS\bar{\alpha}} \quad (17)$$

Comparing Eq. (17) to (4), parameter $\lambda$ in the revised theory can be derived as $cS\bar{\alpha}/4V(1-e^{-1}\bar{\alpha})$. Using parameter $\lambda$, the reverberation decay from the steady state can be expressed using Eq. (18), where $R$ is the room constant and $R = S\bar{\alpha}/(1-\bar{\alpha})$.

$$E(t) = \frac{W}{V}\frac{4V(1-e^{-1}\bar{\alpha})}{cS\bar{\alpha}}\exp\left[-\frac{cS\bar{\alpha}}{4V(1-e^{-1}\bar{\alpha})}t\right]$$
$$= \frac{4W}{c}\left(\frac{1}{A}-\frac{1}{eS}\right)\exp\left[-\left(\frac{1}{A}-\frac{1}{eS}\right)^{-1}\frac{ct}{4V}\right]$$
$$= \frac{4W}{c}\left(\frac{1}{R}+\frac{1-e^{-1}}{S}\right)\exp\left[-\left(\frac{1}{R}+\frac{1-e^{-1}}{S}\right)^{-1}\frac{ct}{4V}\right]$$
(18)

### 4.2. Reverberation time and average sound pressure level

From Eq. (7), the reverberation time in the revised theory can be defined using Eq. (19).

$$T = \frac{0.163V}{A}(1-e^{-1}\bar{\alpha})$$
$$= 0.163V\left(\frac{1}{A}-\frac{1}{eS}\right)$$
$$= 0.163V\left(\frac{1}{R}+\frac{1-e^{-1}}{S}\right)\ [s] \quad (19)$$

The reverberation time from Eq. (19) is equal to Sabine's reverberation time multiplied by $(1-e^{-1}\bar{\alpha})$. Stephenson [19] clarified that for small $\bar{\alpha}$, Eyring's reverberation time is nearly equal to Sabine's reverberation time multiplied by $(1-0.5\bar{\alpha})$. Therefore, because of the factor $(1-e^{-1}\bar{\alpha})$ of the revised theory, the reverberation time obtained using the revised theory would be shorter than that of Sabine's theory and longer than that of Eyring's theory.

Substituting $\lambda = cS\bar{\alpha}/4V(1-e^{-1}\bar{\alpha})$ into Eq. (8), the average sound pressure level $L_p$ in the revised theory can be expressed using Eq. (20). According to Eq. (20), $L_p$ includes the components of the reflected $4/R$ and direct $(1-e^{-1})/S$ sounds.

$$L_p = L_w + 10\log_{10}\left(\frac{1}{A}-\frac{1}{eS}\right)+6 \quad [dB]$$
$$= L_w + 10\log_{10}\left(\frac{1}{R}+\frac{1-e^{-1}}{S}\right)+6 \quad [dB] \quad (20)$$

### 4.3. Mean free path of direct sound

When the mean free path of direct sound is $\bar{\ell}_d$, the energy density of the direct sound $E_d$ can be expressed as $(W/V)(\bar{\ell}_d/c)$. Comparing this to Eq. (16), the mean free path of direct sound $\bar{\ell}_d$ can be expressed as Eq. (21).

$$\bar{\ell}_d = \bar{\ell}(1-e^{-1}) = \frac{4V}{S}(1-e^{-1}). \quad (21)$$

In the revised theory, the mean free paths of direct sound $\bar{\ell}_d$ and reflected sound $\bar{\ell}$ differ, as demonstrated in Eq. (21). The mean free path of direct sound $\bar{\ell}_d$ is shortened by just the length of $\bar{\ell}/e$ compared to that of reflected sound $\bar{\ell}$.

### 4.4. Interpretation of parameter $\lambda$ in reverberation decay

A new concept of mean absorption free path $\bar{\ell}_a$, which differs from the mean free path is introduced here. Minimum sound energy which cannot divide any more is assumed as a sound particle. The particle travels with the speed of sound in a room. When the particle hits a wall surface with absorption coefficient $\alpha$, the particle either perfectly absorbed in probability $\alpha$ or perfectly reflected in probability $(1-\alpha)$. The mean absorption free path can be defined as an expected value of distance where the particle travels in space until absorbed. Propagation of sound energy can be represented as propagation of numerous particles. Because diffused sound field can be considered to have ergodic property, the mean absorption free path of one particle equals to that of the aggregation of the particles.

Sabine's reverberation decay by Eq. (9) can be also expressed using the mean free path $\bar{\ell}$ as Eq. (22).

$$E(t) = \frac{W}{cV}\cdot\frac{\bar{\ell}}{\bar{\alpha}}\exp\left[-\left(\frac{\bar{\ell}}{\bar{\alpha}}\right)^{-1}ct\right]. \quad (22)$$

The factor $\bar{\ell}/\bar{\alpha}$ can be interpreted as the mean absorption free path in Sabine's theory, $\bar{\ell}_{a\_Sab}$.

The reverberation decay of the revised theory by Eq. (18) can be transformed using the mean free path $\bar{\ell}$ as Eq. (23).

Revision of Sabine's Reverberation Theory

$$E(t) = \frac{W}{cV}\left(\frac{\bar{\ell}}{\bar{\alpha}} - \frac{\bar{\ell}}{e}\right)\exp\left[-\left(\frac{\bar{\ell}}{\bar{\alpha}} - \frac{\bar{\ell}}{e}\right)^{-1}ct\right]. \quad (23)$$

From Eq. (23), parameter $\lambda$ in the revised theory can be expressed also as $c/(\bar{\ell}/\bar{\alpha} - \bar{\ell}/e)$. The factor $(\bar{\ell}/\bar{\alpha} - \bar{\ell}/e)$ means the mean absorption free path in the revised theory, $\overline{\ell_{a\_Rev}}$. Similar to the mean free path of direct sound $\overline{\ell_d}$, the mean absorption free path in the revised theory $\overline{\ell_{a\_Rev}}$ is shortened by just the length of $\bar{\ell}/e$ regardless of $\bar{\alpha}$ compared to that of Sabine's theory, $\overline{\ell_{a\_Sab}}$. We can easily understand this by what the revised theory distinguishes between the mean free paths of direct sound $\overline{\ell_d} = \bar{\ell}(1 - e^{-1})$ and reflected sound $\bar{\ell}$.

Here, we examine the reason why the length of $\bar{\ell}/e$, which seems to be related to direct sound, influences the reverberation process far beyond time $t = \bar{\ell}/c$, which should not relate to the direct sound.

As mentioned above, the revised theory assumed that the perfect diffusion state maintains both in the steady state and in the entire reverberation process. Therefore, the sound fields in the steady state and throughout the reverberation process cannot be distinguished, except for the sound energy density. Both are the same statistically. Thus, the energy decay from $t=0$ and that from any time even far beyond time $t = \bar{\ell}/c$ essentially should have the same decay mechanism. Based on this, we can understand that both $\overline{\ell_d}$ and $\overline{\ell_{a\_Rev}}$ are shortened by just the same length $\bar{\ell}/e$ in the revised theory.

## 5. COMPARISON OF THEORIES WITH COMPUTER SIMULATION

### 5.1. Simulation method

To verify the revised theory, simulations were performed using the sound ray tracing method. The mean free paths of direct and reflected sound, the mean absorption free path, the reverberation time and the average sound pressure level were calculated from the simulations to compare with the values calculated using each theory.

Travelling sound particles (representing the sound energy) were used as the sound rays to simulate the propagation of the sound energy. Two kinds of the sound ray tracing methods were employed in this study. One is the method (Method-1) in which the energy of the sound particle was multiplied by (1-α) according to the absorption coefficient α of the wall surface at each reflection. The other is the method (Method-2) in which each sound particle was completely absorbed with the probability α or completely reflected with probability (1-α) at each reflection. The number of sound particles, $N$, decreases with each reflection in Method-2, but not in Method-1. In Method-2, the decrease in the number of sound particles corresponds to the decrease in the sound energy density in a room. The mean free path of direct sound and the mean absorption free path can be obtained directly by using Method-2. However, the dynamic range $D$ of the reverberation decay obtained by Method-2 is limited as $D < 10\log_{10}N$ depending on the particle numbers $N$ used in the simulation. Thus, Method-1 was used to obtain the reverberation time and the average sound pressure level, and Method-2 was used to obtain the mean free path of direct sound and the mean absorption free path.

Two rectangular rooms with dimensions of 15 m × 10 m × 5 m and 10 m × 10 m × 10 m were used for the simulation. Absorption coefficients were set from 0.2 to 0.8 at every step of 0.1. The absorption coefficients were uniform across all walls.

The revised theory is constructed based on the average impulse response described in the section 2.1. Thus, the average impulse response is simulated first, and then the reverberation decay is obtained by performing the Schroeder integration on the average impulse response. A large number of point sources must be uniformly distributed in order to simulate a perfect diffusion state as the initial state of the sound field. Due to limited computer resources and time, it is not possible to use these conditions. Instead, to simulate numerous point sources, $10^8$ sound particles were randomly distributed throughout the room. The traveling direction of each particle was determined randomly using uniform spherical random numbers. This is statistically equivalent to the uniform distribution of an infinite number of point sound sources in a room. The total sound power $W$ of all particles was set to 1.0 [watt]. The time step $\Delta t$ was $0.01/c$ because the travel of sound particles is calculated every 0.01 m.

In Method-1, information such as energy, direction of propagation, number of times reflected, of every particle were recorded in every calculation step. In Method-2, information such as being reflected or absorbed, distance of propagation, direction of propagation, number of times reflected, of every particle were recorded in every calculation step. The mean free paths of direct and reflected sound and the mean absorption free path were calculated using the information.

When each sound particle is reflected on the wall, it is diffusely reflected according to Lambert's cosine law. From the total energy of all the particles at each time step $\Delta t$, the change in the sound energy density $\Delta E$ over time, namely the average impulse response, was calculated. And then the reverberation decay curve $E(t)$ was calculated by performing the Schroeder integration on the average impulse response. The calculated $E(t)$ means the energy decay from the steady state. And because of the simulated numerous point sound sources and the diffuse reflections according to Lambert's cosine law, the diffusion state can be approximately maintained in the steady state and reverberation process.

The reverberation time was calculated from the -5 dB to -35 dB slope of the reverberation decay curve. The average sound pressure level was obtained from the initial energy of the reverberation decay $E(0)$.

### 5.2. Results and discussion

Fig. 2 shows the comparison between simulation methods Method-1 and Method-2 in reverberation decay curves with average absorption coefficients of 0.2 and 0.5. As a result, both Method-1 and Method-2 agreed within the limit of the dynamic range $D$.

Fig. 3 shows comparisons between theories and simulation in the mean free paths of direct sound (mfp_dir_Revised) and reflected sound (mfp_ref), and in the mean absorption free path of Sabine's theory (mfp_abs_Sabine) and the revised theory (mfp_abs_Revised). Looking at the simulation results of mfp_dirs and mfp_refs first, the mfp_dirs were shortened by the length of about $\bar{\ell}/e$ compared to the mfp_refs, regardless of the absorption coefficients.

These simulation results were in better agreement with the results of mfp_dir_Revised derived from the revised theory. Next, looking at the simulation results of the mean absorption free path mfp_abs, the mfp_abs decreased as the average absorption coefficient increased. These simulation results were in better agreement with the theoretical values of mfp_abs_Revised rather than mfp_abs_Sabine. The mfp_abss were shortened by the length of about $\bar{\ell}/e$ compared to the mfp_abs_Sabine. This phenomenon is also one of the consequences of the revised theory.

Fig. 4 compares the theoretical solutions with the simulation results in the reverberation times and average sound pressure levels. The results of the simulations agreed better with the revised theory rather than Sabine's and Eyring's theories. The reverberation time obtained using the revised theory was shorter than that of Sabine's theory and longer than that of Eyring's theory. The average sound pressure levels were lower than those of existing theories.

Fig. 5 compares reverberation decay curves obtained by the theories with the simulation results in cases of $\alpha = 0.2$ and $\alpha = 0.5$. The reverberation decay curves of the simulations agreed better with the revised theory rather than Sabine's and Eyring's theories. Sabine's decay curves were always overestimated and Eyring's ones were always underestimated.

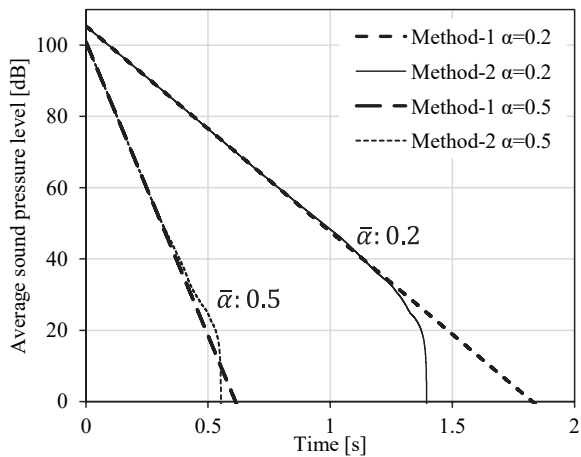
Rectangular room with a dimension of 15 m×10 m×5 m

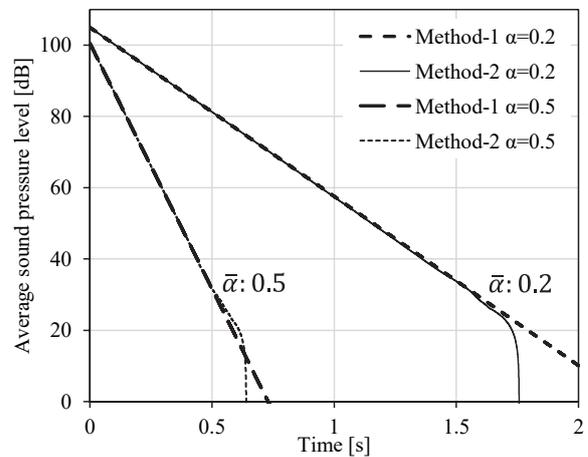
Rectangular room with a dimension of 10 m×10 m×10 m

**Fig. 2** Comparison between simulation methods Method-1 and Method-2 in reverberation decay curves

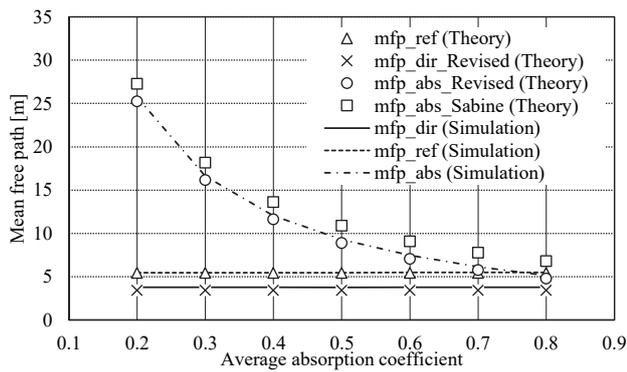
Rectangular room with a dimension of 15 m×10 m×5 m

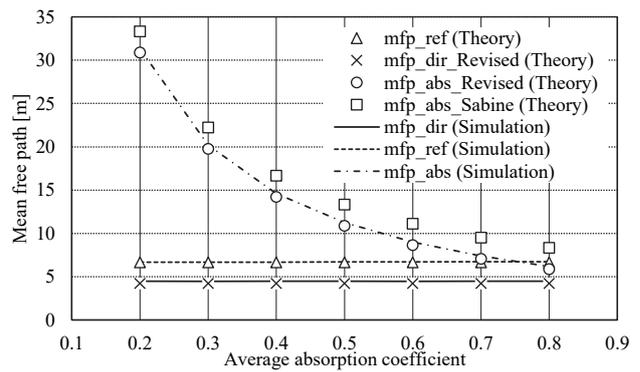
Rectangular room with a dimension of 10 m×10 m×10 m

**Fig. 3** Comparison between theories and simulation in mean free paths

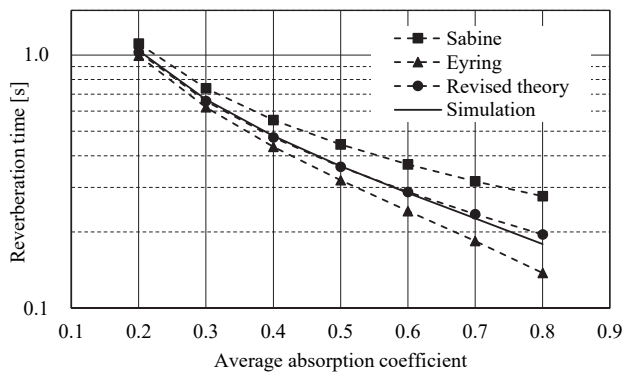

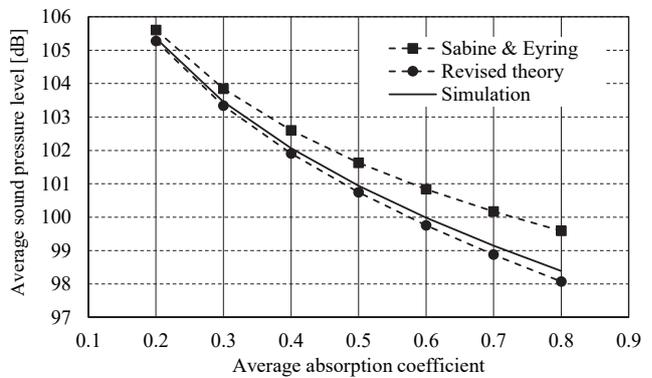

Rectangular room with a dimension of 15 m×10 m×5 m

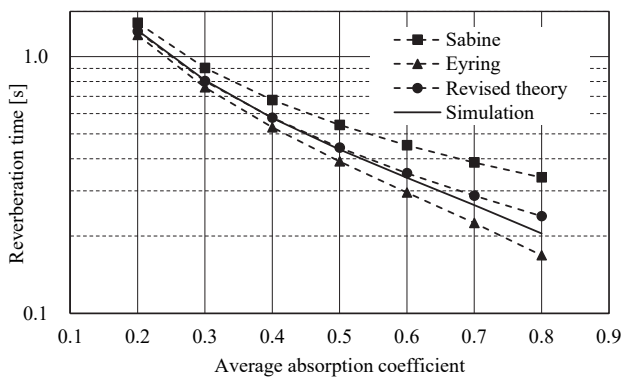

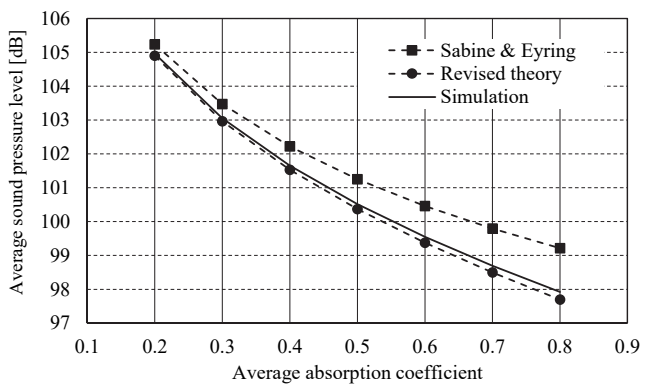

Rectangular room with a dimension of 10 m×10 m×10 m

**Fig. 4** Comparison between theories and simulation (left: reverberation time, right: average sound pressure level).

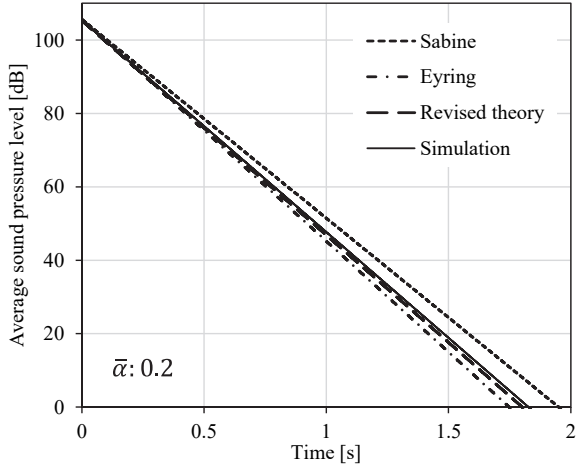
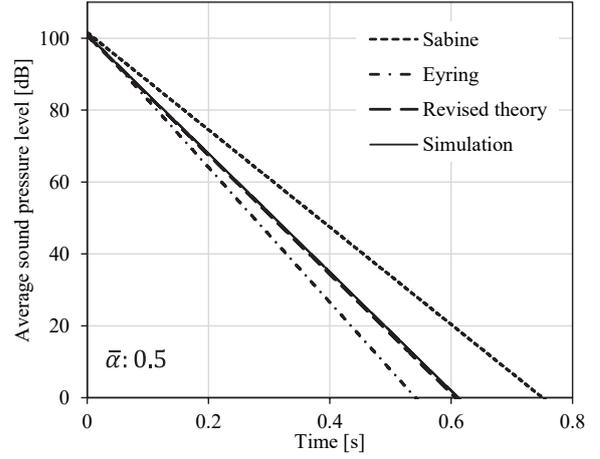

Rectangular room with a dimension of 15 m × 10 m × 5 m

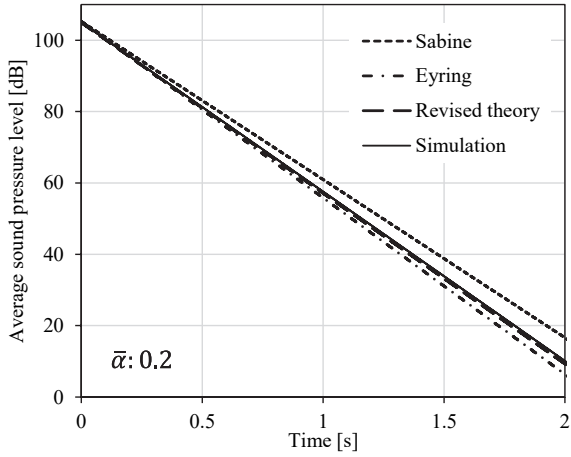
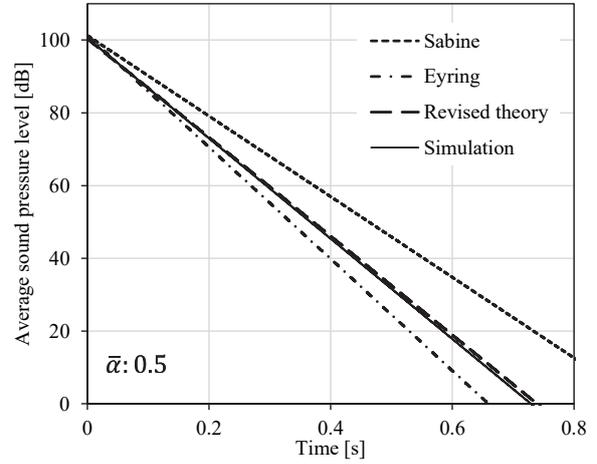

Rectangular room with a dimension of 10 m × 10 m × 10 m

**Fig. 5** Comparison between theories and simulation in reverberation decay curves (left: $\bar{\alpha} = 0.2$, right: $\bar{\alpha} = 0.5$).

## 6. CONCLUSIONS

In this study, a theoretical framework for the consistent reverberation theory was presented. Based on this theoretical framework, Sabine's reverberation theory was revised by following an approach that is different from Eyring's. The reverberation time obtained using the revised theory is equal to Sabine's reverberation time multiplied by $(1 - e^{-1}\bar{\alpha})$. The newly defined reverberation time was shorter than that obtained using Sabine's theory and longer than that obtained using Eyring's theory.

The computer simulation results, obtained using the ray-tracing methods in rooms with walls, which are totally diffuse and follow Lambert's cosine law, confirmed that the revised theory explains the results more reasonably than Sabine's and Eyring's theories in the practical range of absorption coefficients from 0.2 to 0.8.

In the future, it will be necessary to verify the revised theory under various sound field conditions.

## REFERENCES

[1] W. C. Sabine: Collected papers on acoustics, Harvard University Press (1927).
[2] C. F. Eyring, "Reverberation time in "dead" rooms", J. Acoust. Soc. Am., 1, pp.217-241 (1930).
[3] G. Millington, "A modified formula for reverberation", J. Acoust. Soc. Am., pp.69-82, 5/Jan. (1932).
[4] V. O. Knudsen, "Absorption of sound in air, in oxygen, and in nitrogen effects of humidity and temperature", J. Acoust. Soc. Am., 5, pp.112-121 (1933).
[5] D. Fizzroy, "Reverberation formula which seems to be more accurate with nonuniform distribution of absorption", J. Acoust. Soc. Am., 31(7), pp.893-897 (1959).
[6] H. Arau-Puchades, "An improved reverberation formula", Acustica, 65(4), pp.163-180 (1988).